\documentclass[preprint,aps,floats,subeqn,showpacs]{revtex4}
\usepackage{epsfig} 

\newcommand{\be}{\begin{equation}}
\newcommand{\ee}{\end{equation}}
\newcommand{\bea}{\begin{eqnarray}}
\newcommand{\eea}{\end{eqnarray}}
\newcommand{\bml}{\begin{mathletters}}
\newcommand{\eml}{\end{mathletters}}

\begin{document}

\tighten
\preprint{IUB-TH-0410}
\draft




\title{Gravity localisation in a 6-dimensional
brane world\footnote{Talk presented at the
6th Alexander Friedmann International Seminar on Gravitation
and Cosmology, Carg\`ese,
Corsica, 28.6.-3.7.2004}   }
\renewcommand{\thefootnote}{\fnsymbol{footnote}}
\author{Betti Hartmann}
\affiliation{School of Engineering and Sciences, International
University Bremen (IUB), 28725 Bremen, Germany; e-mail: 
b.hartmann@iu-bremen.de}
\author{Yves Brihaye }
\affiliation{ 
Facult\'e des Sciences, Universit\'e des Mons-Hainaut,
7000 Mons, Belgium; e-mail: yves.brihaye@umh.ac.be}
\date{\today}
\setlength{\footnotesep}{0.5\footnotesep}

\begin{abstract}
We study a 6-dimensional Einstein-Born-Infeld-Higgs model.
In the limit of infinite Born-Infeld coupling, this model
reduces to an Einstein-Abelian-Higgs model, in which gravity localising
solutions were shown to exist. In this proceeding, we discuss
further properties of the gravity localising solutions as well
as of the solutions in the limit of vanishing cosmological constant.

\end{abstract}
\maketitle
\section{Introduction}
Recently, brane world models with large (non-) compact extra dimensions
have gained a lot of interest \cite{dvali,anton,arkani,rs1,rs2}.
These models assume that we live on a 3-brane embedded in a higher dimensional
manifold. The standard model fields and particles are confined
to the brane, while gravity, which is a property of space-time itself,
lives in the full dimensions. The idea that matter is confined to a lower
dimensional manifold is not a new idea. The localisation of fermions
on a domain wall has been discussed in \cite{ruba}. Recently, it was newly motivated
by results from string theory. In type I string theory so-called Dp-branes
exist on which open strings, which represent matter fields, end.
Gravitational fields, which are represented by closed strings, live in the full
dimensions. However, as is well known, Newton's law in 4 dimensions
is well tested down to $0.2$ mm. Thus appropriate brane world models
should localise gravity ``well enough'' to the 3-brane.

In \cite{shapo}, the localisation of gravity in more than one extra dimension
has been discussed. This was achieved by studying higher dimensional
topological defects such as Nielsen-Olesen strings \cite{shapo1,shapo2}
and magnetic monopoles
\cite{shapo3} in 6 and 7 space-time dimensions, respectively.
It was found \cite{shapo,shapo1} that gravity-localising (so-called ``warped'')
solutions are possible if certain relations between the defect's tensions
hold.

Originally introduced to remove singularities associated with point-like
charges in electrodynamics \cite{BI}, the generalisation
of the Born-Infeld (BI) action to non-abelian gauge fields
has gained a lot of interest in topics
related to string theory \cite{pol,tse}. It became apparent
that when studying low energy effective actions
of string theory  the part of the Lagrangian containing
the abelian Maxwell field strength tensor and its non-abelian 
counterpart in Yang-Mills theories has to be replaced by a corresponding
(resp. abelian and non-abelian) BI term. That's why it seems
interesting to generalise the brane world scenario 
for 6-dimensional Nielsen-Olesen strings recently proposed by Giovannini, 
Meyer
and Shaposhnikov (GMS) \cite{shapo2} to Born-Infeld actions.
This was done in \cite{bh} and it was found that gravity-localising 
solutions don't exist in a 6-dimensional Einstein-Born-Infeld-Higgs model.
Here, we review the results and extend the discussions done in \cite{bh}.

\section{The model}
The action reads:
\begin{eqnarray}
S&=&S_{gravity}+S_{brane}=-\int d^6 x \sqrt{-g} \frac{1}{16\pi G_6} \left(R+2\hat{\Lambda}_6\right) \nonumber \\
&+& \int d^6x \sqrt{-g} \left(\hat{\beta}^2(1-{\cal R}) +\frac{1}{2} D_M\phi D^M \phi^* -\frac{\lambda}{4}(\phi^*\phi-v^2)^2\right)
\end{eqnarray}
where $\hat{\Lambda}_6$ is the bulk cosmological constant, $G_6$ is the fundamental
gravity scale with $G_6=1/M^4_{pl(6)}$ and $g$ the determinant
of the 6-dimensional metric. Moreover, we have the covariant derivative
$D_M=\nabla_M-ie A_M$ and the field strength $F_{MN}=\partial_M A_N-\partial_N A_M$ of the U(1)
gauge potential $A_M$. $v$ is the vacuum expectation
value of the complex valued Higgs field $\phi$ and $\lambda$ is the self-coupling
constant of the Higgs field.

The Ansatz reads \cite{shapo1,shapo2,bh}:
\begin{equation}
ds^2=M^2(r)\left(dx_1^2-dx_2^2-dx_3^2-dx_4^2\right)-dr^2-l^2(r)d\theta^2
\end{equation}
for the metric and
\begin{equation}
\phi(r,\theta)=vf(r)e^{in\theta} \ \ , 
\ \ A_{\theta}(r,\theta)=\frac{1}{e}\left(n-P(r)\right)
\end{equation}
for the gauge and the Higgs field \cite{no}, where $n$ is the vorticity
of the string. The equations of motion can be computed easily (see \cite{bh} for details)
and depend after the rescalings $x=\sqrt{\lambda} v r$, $L(x)=\sqrt{\lambda} v l(r)$ only 
on the coupling constants $\alpha=e^2/\lambda$, $\gamma^2=8\pi G_6 v^2$, $\Lambda=\hat{\Lambda}_6/(\lambda v^2)$,
$\beta^2=\hat{\beta}^2/(\lambda v^2)$.

\section{Numerical results}
The system of equations has been solved numerically subject to the following
boundary conditions:
\begin{equation}
f(0)=0 \ , \ P(0)=n \ , \ M(0)=1 \ , \ M'|_{x=0}=0 \ , \ L(0)=0 \ , \ L'|_{x=0}=1
\end{equation}
at the origin and
\begin{equation}
f(\infty)=1 \ \ , \ \ P(\infty)=0
\end{equation}
at infinity.

\begin{figure}[htb]
\centering
\epsfysize=10cm
\mbox{\epsffile{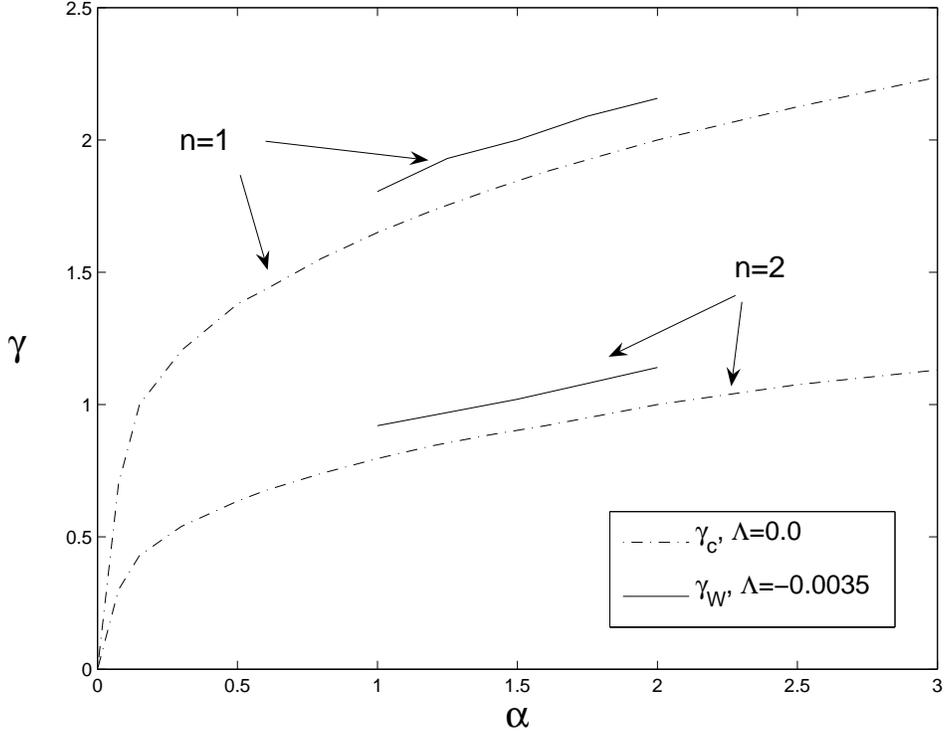}}
\caption{\label{fig1} 
The value $\gamma_c$ for $\Lambda=0$ as well as
$\gamma_W$ for $\Lambda_W=-0.0035$ are shown in dependence on 
$\alpha$ for $\beta^2=\infty$ and  $n=1$, $n=2$.}
\end{figure}

We have studied the set of equations numerically. We find that several 
branches
of solutions exist. The pattern of these branches is very similar than in 
the
4 dimensional case \cite{gstring}. For small gravitational
coupling, $\gamma < \gamma_{c}(\alpha,\Lambda)$, the gravitational
deformations of the Nielsen-Olesen string solutions as well as so-called 
Melvin 
solutions
exist for the same values of the coupling constants. They differ, however,
in their asymptotic behaviour. The metric functions on the
string branch behave like $M(x\rightarrow \infty)\rightarrow a$,
$L(x\rightarrow\infty)\rightarrow bx+c$, where $a$, $b$, $c$ are constants
depending on $\gamma$, $\alpha$ and $\Lambda$.
The metric functions on the Melvin branch have the following asymptotic 
behaviour:
$M(x\rightarrow \infty)\rightarrow Ax^{2/5}$,
$L(x\rightarrow\infty)\rightarrow Bx^{-3/5}$, where again $A$ and $B$ are 
constants
depending on the coupling constants. For $\gamma > 
\gamma_{c}(\alpha,\Lambda)$
closed solutions with zeros of the metric functions exist.
For the so-called inverted string branch, which represents the
strong gravity analogue of the string branch, the metric function $L(x)$ 
possesses
a zero, $L(x=x_0)=0$, while $M(x)$ stays finite at this $x_0$.
Furthermore, a so-called Kasner branch exists. The solutions on this
branch possess a zero of the metric function $M(x)$, $M(x=\tilde{x}_0)=0$, 
while
$L(x\rightarrow \tilde{x}_0)\rightarrow \infty$. In Fig. \ref{fig1},
we show the dependence of the critical value $\gamma_{c}$ on $\alpha$
for $\Lambda=0$ and $n=1$, $2$. 
As is clearly seen, $\gamma_{c}$ increases with increasing $\alpha$
and tends to zero for $\alpha\rightarrow 0$. This likely stems from the 
fact
that the mass of the vortex decreases with increasing
$\alpha$ and thus the strong gravity limit
is reached for higher values of the gravitational coupling.
Furthermore, 
the 
critical
value of $\gamma$ for a fixed $\alpha$ decreases with the winding number
$n$ as can be seen when comparing the two curves for $n=1$ and $n=2$.
Again, this can be explained by the increased mass
of the vortex solution.

Interestingly, the pattern of solutions persists for $\Lambda < 0$.
For the $\beta^2=\infty$ limit, it was found in \cite{shapo1} that 
so-called
``anti-warped'' and gravity-localising (``warped'') solutions
exist. However, no classification was done. In \cite{bh}, it was 
demonstrated
that the ``anti-warped'' solutions are the by the presence of the 
cosmological 
constant
deformed string and Melvin solutions, while the warped
solutions are the solutions on the limiting line between the
Kasner and inverted string solutions.
Of course, this is only possible if a fine-tuning between the
coupling constants is done, i.e. warped solutions exist only
for specific triplets $(\gamma_{W},\alpha_{W},\Lambda_{W})$
of the coupling constants.
In Fig.\ref{fig1}, we give the value of $\gamma_{W}$ in dependence on 
$\alpha$ for
$\Lambda_{W}=-0.0035$ and $n=1$, $n=2$.

As can be seen from Fig.\ref{fig1} 
gravity-localising solutions for a fixed $\alpha_W$, $\Lambda_W$
exist for lower values of the gravitational coupling for the $n=2$
solutions than for the $n=1$ solutions.

It is interesting to study how the value $\gamma_{W}$ for $\Lambda < 0$ is 
related
to the value $\gamma_c$ for $\Lambda=0$ for a fixed value of $\alpha$.
We have studied this question in some detail and as can be seen from 
Fig.\ref{fig1}, our results reveal that the curves are more or less 
parallel. This holds true for both $n=1$ and $n=2$. We thus find the
following approximate relation:
\begin{equation}
\gamma_W(\alpha,\Lambda,n)\approx  
\gamma_c(\alpha,\Lambda=0,n)+3\vert\Lambda\vert^{1/2}
\end{equation}

This completes the analysis done in \cite{shapo1,bh}
in the sense that here we have determined the dependence of $\gamma_c$ 
and $\gamma_W$ on both $\alpha$ and $n$.

Let us remark that it was 
found \cite{bh} by an analytic argument that no gravity-localising
solutions exist if $\beta^2 < \infty$.
Of course, this result could be different if higher order corrections
(e.g. higher derivative terms of the gauge field), which appear in the low 
energy
effective action of string theory, are included.

\section{Conclusions}
Brane world models have attracted a lot of attention recently.
Since Newton's law is very well tested  in 4 dimensions now, proper models
should localise gravity well enough to the 3-brane.
Randall-Sundrum models possess one extra codimension. For more than one
extra dimension, alternative models have to be looked for. In 
\cite{shapo,shapo1,shapo2,shapo3,bh},
models with 2 and 3 codimensions have been studied by using topological
defects in higher dimensions.
Here, we have discussed further properties of the solutions
in the 6-dimensional Einstein-Abelian-Higgs model. We have determined
the $\alpha$ dependence of the critical gravitational coupling $\gamma_c$ 
for 
$\Lambda=0$.  Further we have determined how the gravitational
coupling $\gamma_W$, for which gravity-localising solutions exist, depends
on $\alpha$ and $n$. We found an interesting relationship between
$\gamma_c$ and $\gamma_W$.

\section*{Acknowledgements}
We thank the organisers of the 6th Alexander Friedmann Seminar and the 
Institut d'\'Etudes Scientifiques de Carg\`ese
for their hospitality.

\end{document}